\documentclass[12pt]{JHEP3}
\usepackage[centertags]{amsmath}
\usepackage{amssymb}
\usepackage{epsfig}
\usepackage{amsmath}
\usepackage{graphicx}
\usepackage{afterpage}
\usepackage{cite}

\def\beq{\begin{equation}}
\def\eeq{\end{equation}}
\def\bea{\begin{eqnarray}}
\def\eea{\end{eqnarray}}
\def\bwt{\begin{widetext}}
\def\ewt{\end{widetext}}
\def\nn{\nonumber}
\def\roughly#1{\mathrel{\raise.3ex\hbox
{$#1$\kern-.75em\lower1ex\hbox{$\sim$}}}}

\def\ks{K_S}
\def\kbar{{\bar K}^0}
\def\bd{B^0_d}
\def\bs{B_s^0}
\def\bdbar{{\bar B}^0_d}
\def\bsbar{{\bar B}_s^0}

\def\btod{{\bar b} \to {\bar d}}
\def\btos{{\bar b} \to {\bar s}}
\def\order{\lower 1.8ex \hbox{\LARGE\~{}}}

\def\bskk{\bs \to K^+ K^-}
\def\bdpipi{\bd \to \pi^+ \pi^-}
\def\bdpik{\bd\to \pi^\mp K^\pm}

\def\fbar{{\bar f}}

\def \bwt{\begin{widetext}}
\def \ewt{\end{widetext}}

\def \bin{{\rm bin}}
\def \DP{{\rm DP}}

\def \bma{\begin{matrix}}
\def \ema{\end{matrix}}

\def \ll{\left|}

\def \lle{\left\langle}
\def \rre{\right\rangle}

\def \<{\left<}
\def \>{\right>}
\def \[{\left[}
\def \]{\right]}
\def \({\left(}
\def \){\right)}

\def \l.{\left.}
\def \r.{\right.}
\def \nn{\nonumber}
\def \nl{\nn\\}

\def \hf{\frac{1}{2}}

\def \Re{{\rm Re}}
\def \Im{{\rm Im}}

\def \ob{\overline{B}^0}
\def \of{\overline{f}}
\def \oA{\overline{A}}
\def \tA{\widetilde{A}}

\def \ola{\overline{\la}}
\def \ox{\overline{x}}

\def \btod{{\bar b}\to{\bar d}}
\def \btos{{\bar b}\to{\bar s}}

\def \cM{{\cal M}}

\def \la{\lambda}

\def \Ga{\Gamma}

\def \De{\Delta}

\def \phys{{\rm phys}}

\newcommand{\ket}[1]{\ll #1 \rre}
\newcommand{\brkt}[2]{\lle #1 | #2\rre}

\title{\boldmath Using U Spin to Extract $\gamma$ from Charmless $B\to PPP$ Decays}

\author{
Bhubanjyoti Bhattacharya and David London
\\
Physique des Particules, Universit\'e de Montr\'eal, \\
C.P. 6128, succ. centre-ville, Montr\'eal, QC, Canada H3C 3J7 \\
E-mail:
\email{bhujyo@lps.umontreal.ca},
\email{london@lps.umontreal.ca}
}

\abstract{Some years ago, a method was proposed for measuring the
  CP-violating phase $\gamma$ using pairs of two-body decays that are
  related by U-spin reflection ($d \leftrightarrow s$). In this paper
  we adapt this method to charmless $B\to PPP$ decays. Time-dependent
  Dalitz-plot analyses of these three-body decays are required for the
  measurement of the mixing-induced CP asymmetries.  However, isobar
  analyses of the decay amplitudes are not necessary.  A potential
  advantage of using three-body decays is that the effects of U-spin
  breaking may be reduced by averaging over the Dalitz plot. This can
  be tested independently using the measurements of direct CP
  asymmetries and branching ratios in three-body charged $B$ decays.}


\preprint{UdeM-GPP-TH-15-240}

\begin{document}

\section{Introduction}

The standard model (SM) explanation of CP violation is that it is due
to a phase in the Cabibbo-Kobayashi-Maskawa (CKM) quark mixing matrix.
This phase information is elegantly encoded in the unitarity triangle,
whose interior CP-violating angles are $\alpha$, $\beta$ and $\gamma$
\cite{pdg}. Using $B$ decays, a great deal of effort has gone into
measuring these angles in many different ways, along with the sides of
the unitarity triangle, to search for inconsistencies that would
indicate the presence of new physics (NP). Unfortunately, to date no
such indications have been seen. This suggests that the NP is more
massive than hoped for (which is consistent with the absence of NP
signals at the LHC), and that the observation of its effects on CP
violation in the $B$ system will require measurements of greater
precision.

One interesting procedure for searching for NP involves the CKM phase
$\gamma$. The conventional way of measuring $\gamma$ uses the
tree-level decay $B^- \to D^{(*)} K^{(*)-}$
\cite{GL,GW,ADS,GGSZ,GGSZBelle}. Its latest value is $\gamma =
(71.7^{+7.1}_{-7.4})^\circ$ \cite{ckmfitter}.  However, suppose that
$\gamma$ could be measured using decays that have significant (gluonic
or electroweak) penguin contributions. If NP is present, it is likely
to affect the (loop-level) penguins, in which case the extracted value
of $\gamma$ would be different from that found using $B^- \to D^{(*)}
K^{(*)-}$.  That is, one can probe NP by comparing the ``tree-level''
and ``loop-level'' values of $\gamma$. (But note that, if there is NP,
the ``loop-level'' value of $\gamma$ will not be constant. It will
generally vary, depending on which decays are used for its
extraction.)

One example of this involves $B \to \pi K$ decays. (In what follows,
we briefly describe the method, but we refer the reader to
Ref.~\cite{BtopiK} for full details.) There are four such decays: $B^+
\to \pi^+ K^0$, $B^+ \to \pi^0 K^+$, $\bd \to \pi^- K^+$ and $\bd \to
\pi^0 K^0$. Using these processes, one can measure nine observables:
four branching ratios, four direct CP asymmetries, and one indirect
(mixing-induced) CP asymmetry. However, assuming flavor SU(3)
symmetry, the amplitudes can be written in terms of eight theoretical
parameters: the magnitudes of the diagrams $P'_{tc}$, $T'$, $C'$,
$P'_{uc}$, three relative strong phases, and the weak phase $\gamma$.
(The value of the weak phase $\beta$ is taken from the measurement of
indirect CP violation in $\bd(t)\to J/\Psi \ks$ \cite{ckmfitter}.)
With more observables than theoretical parameters, one can perform a
fit to extract $\gamma$. The value found is $\gamma = (35.3 \pm
7.1)^\circ$ \cite{BtopiK}, which differs from the tree-level value of
$\gamma$ by $3.5\sigma$. While this is intriguing, one must remember
that there is also an unknown theoretical uncertainty due to SU(3)
breaking. Before any conclusions can be drawn, there must be other,
independent determinations of loop-level values of $\gamma$.

In 1999, R. Fleischer proposed a method for extracting $\gamma$ from
$\bskk$ and $\bdpipi$, two decays whose amplitudes are related by
U-spin ($d\leftrightarrow s$) symmetry \cite{Fleischer99}. Since
penguin contributions are important for such decays, this method would
determine a loop-level value of $\gamma$. It requires the measurement
of the branching ratios and CP asymmetries, both direct and indirect,
of both decays.  This method is unaffected by final-state
interactions; its theoretical accuracy is limited only by the size of
U-spin-breaking effects. The factorizable U-spin-breaking corrections
are calculable theoretically in terms of form factors and decay
constants \cite{Fleischer99,Fleischer1,Fleischer2}.  However, the
precise value of the nonfactorizable U-spin-breaking correction is
unknown, though it may be sizeable \cite{Beneke}.

Recently, the direct and indirect CP asymmetries in $\bskk$ were
measured by the LHCb Collaboration \cite{bskkLHCb}, and they carried
out the above extraction of $\gamma$ \cite{LHCbUspin}. Allowing for a
U-spin-breaking error of 50\%, they find
$\gamma=(63.5^{+7.2}_{-6.7})^\circ$. However, if the theoretical error
is $\ge 60\%$, the uncertainty on $\gamma$ is much larger.

It was pointed out in Ref.~\cite{Fleischer99} that, with an additional
dynamical assumption, one could replace $\bskk$ with $\bdpik$, and
analyses with this second decay were carried out in
Refs.~\cite{Fleischer1,Fleischer2,Silva}.  However,
Ref.~\cite{MaxDLUspin} finds that the experimental data suggest that
there may be a large nonfactorizable U-spin-breaking correction
between $\bdpik$ and $\bdpipi$. This would lead to a large (unknown)
theoretical error in the extraction of $\gamma$ using $\bdpik$ and
$\bdpipi$.

The main purpose of the present paper is to note that the method of
Ref.~\cite{Fleischer99} can also be applied to charmless $B \to PPP$
decays ($P$ is a pseudoscalar meson) whose amplitudes are related by U
spin.  The key point is that, by using the Dalitz plots of the
three-body decays, the effect of U-spin breaking may be greatly
reduced. If this is possible -- and there is an independent test to
see if the procedure works -- then the loop-level value of $\gamma$
can be determined with little theoretical error. This will then
provide a clean test for NP.

Note that, under flavor SU(3) symmetry, the three final-state
particles in charmless $B\to PPP$ decays are treated as identical, so
that the six permutations of these particles must be considered. There
have been a number of papers recently that use the fully-symmetric
final state \cite{BPPPdiag,BPPP2,BPPP3,BPPPgamma,fullysym}, which can
be obtained using an isobar analysis of the Dalitz plot. However, we
stress that such an analysis is {\it not} needed for the above method
of extracting $\gamma$ -- the full Dalitz plot is used.

Examples of pairs of decays to which this method can be applied are
(i) $\bs \to \ks \pi^+ \pi^-$ ($\btod$) and $\bd \to \ks K^+ K^-$
($\btos$), and (ii) $\bs \to \ks K^+ K^-$ ($\btod$) and $\bd \to \ks
\pi^+ \pi^-$ ($\btos$). The time-dependent Dalitz plots for $\bd \to
\ks K^+ K^-$ and $\bd \to \ks \pi^+ \pi^-$ were measured by the BaBar
and Belle Collaborations
\cite{KsKKBelle,KsKKBaBar,KspipiBelle,KspipiBaBar}, and a study of
$B^0_{(s)} \to \ks h^+ h^{\prime -}$ was made by the LHCb
Collaboration \cite{Ksh+h'-}. For the $\bs$ decays, it appears that
$\bs \to \ks \pi^+ \pi^-$ is more promising experimentally.  The first
observation of this decay was reported in Ref.~\cite{Ksh+h'-}, and a
study of the future prospects for the measurement of its
time-dependent Dalitz plot was presented in
Ref.~\cite{Gershonetal}. Hopefully the method will be applied to
decays $\bs \to \ks \pi^+ \pi^-$ and $\bd \to \ks K^+ K^-$ to extract
$\gamma$.

In Sec.~2, we briefly discuss Dalitz plots and the distinction between
the final states $f$ and $\fbar$. The U-spin relation between $\btod$
and $\btos$ decays is discussed in Sec.~3. In Sec.~4, we present the
method for extracting $\gamma$ from a Dalitz-plot analysis of
three-body decays. The subject of U-spin-breaking effects -- the
theoretical idea of how they may be reduced in three-body decays, and
experimental tests of this hypothesis -- is examined in Sec.~5. We
conclude in Sec.~6.

\section{Dalitz Plots}
\label{DP}

Three-body $B$ decays are usually described using a Dalitz plot.
Consider the decay $B \to P_1 P_2 P_3$, in which each pseudoscalar
$P_i$ has momenta $p_i$. One can construct the three Mandelstam
variables $s_{ij} \equiv \left( p_i + p_j \right)^2$, where $p_i$ is
the momentum of each $P_i$. These are not independent, but obey
$s_{12} + s_{13} + s_{23} = m_B^2 + m_1^2 + m_2^2 + m_3^2$. The $B \to
P_1 P_2 P_3$ Dalitz plot is a measure of the decay rate as a function
of two Mandelstam variables.

In the present paper we focus on the decays $B_{d,s}^0 \to \ks(p_1)
h^+(p_2) h^-(p_3)$ ($h = K,\pi$). At the quark level, the final states
$f = \ks \pi^+ \pi^-$ and $\ks K^+ K^-$ are self-conjugate. However,
when the momenta are considered, one has $\fbar \ne f$. The point is
that the CP conjugate of $f = \ks(p_1) h^+(p_2) h^-(p_3)$ is $\fbar =
\ks({\bar p}_1)h^-({\bar p}_2)h^+({\bar p}_3)$, where ${\bar p}_i$ is
$p_i$ with the direction of the three-momentum reversed.  Note that
reversing the direction of the three momenta does not affect the
Mandelstam variables, since $s_{ij} = \left( p_i + p_j \right)^2 =
\left({\bar p}_i + {\bar p}_j \right)^2 = {\bar s}_{ij}$. Thus, in
this case the difference between $f$ and $\fbar$ arises from an
exchange of the indices $2$ and $3$.

The distinction between $f$ and $\fbar$ must be kept in mind
throughout the paper. Because $f$ is self-conjugate at the quark
level, both $B^0$ and ${\bar B}^0$ can decay to it, and similarly for
$\fbar$.  Now, at different points in the analysis we consider the
direct CP asymmetry.  However, because $\fbar \ne f$, there are two of
these. One compares $B^0 \to f$ and ${\bar B}^0 \to \fbar$ decays, the
other $B^0 \to \fbar$ and ${\bar B}^0 \to f$. Things are similar for
the indirect CP asymmetry, which arises because both $B^0$ and ${\bar
  B}^0$ can decay to the same final state. Thus, one indirect
asymmetry involves the interference of the amplitudes for $B^0 \to f$
and ${\bar B}^0 \to f$, while the other involves the interference of
${\cal A}(B^0 \to \fbar)$ and ${\cal A}({\bar B}^0 \to \fbar)$.

\section{U-Spin Relation}
\label{Uspinrelsec}

In this section we discuss the U-spin relation that is central to our
method for extracting $\gamma$. We begin by reviewing the relation for
two-body decays.

\subsection{\bf Two-body decays}

Consider a pair of $B \to PP$ decays whose amplitudes are related by
U-spin reflection ($d \leftrightarrow s$). (This discussion follows
Ref.~\cite{Uspinrel}.) One is a $\btod$ decay, the other $\btos$.
There are five such pairs \cite{MaxDLUspin}: ($\bd \to \pi^+ \pi^-$,
$\bs \to K^+ K^-$), ($\bs \to \pi^+ K^-$, $\bd \to \pi^- K^+$), ($B^+
\to K^+ \bar K^0$, $B^+ \to \pi^+ K^0$), ($\bd \to K^0 \kbar$, $\bs
\to \kbar K^0 $), ($\bd \to K^+ K^-$, $\bs \to \pi^+ \pi^-$).

The $\btod$ amplitude can be written
\bea
{\cal A}_d &=& A_u V^*_{ub}V_{ud} + A_c V^*_{cb}V_{cd} + A_t V^*_{tb}V_{td} \nn\\
    &=& (A_u - A_t) V^*_{ub}V_{ud} + (A_c - A_t) V^*_{cb}V_{cd} \nn\\
    &\equiv& V^*_{ub}V_{ud} T_d + V^*_{cb}V_{cd} P_d ~.
\label{Addef}
\eea
In the above, the $A_i$ each represent a linear combination of
diagrams, and we have used the unitarity of the CKM matrix
($V^*_{ub}V_{ud} + V^*_{cb}V_{cd} + V^*_{tb}V_{td} =0$) to write the
second line. $T_d$ and $P_d$ are simply the quantities that are
multiplied by the given CKM matrix elements -- they do not represent
individual ``tree'' and ``penguin'' diagrams. The $\btos$ amplitude
can be written similarly:
\beq
{\cal A}_s = V^*_{ub}V_{us} T_s + V^*_{cb}V_{cs} P_s ~.
\label{Asdef}
\eeq

The CP-conjugate amplitudes ${\bar{\cal A}}_d$ and ${\bar{\cal A}}_s$
are obtained from the above by changing the signs of the weak phases:
\beq
{\bar{\cal A}}_d = V_{ub}V^*_{ud} T_d + V_{cb}V^*_{cd} P_d ~~,~~~~
{\bar{\cal A}}_s = V_{ub}V^*_{us} T_s + V_{cb}V^*_{cs} P_s ~.
\label{2bodyCPconj}
\eeq
We then have
\bea
| {\cal A}_d |^2 - | {\bar{\cal A}}_d |^2 & = & 4 \, {\rm Im} ( V^*_{ub}V_{ud} V_{cb}V^*_{cd} ) \, {\rm Im} (T_d P_d^* ) ~, \nn\\
| {\cal A}_s |^2 - | {\bar{\cal A}}_s |^2 & = & 4 \, {\rm Im} ( V^*_{ub}V_{us} V_{cb}V^*_{cs} ) \, {\rm Im} (T_s P_s^* ) ~.
\eea
Now, the unitarity of the CKM matrix implies \cite{Jarlskog}
\beq
{\rm Im} ( V^*_{ub}V_{us} V_{cb}V^*_{cs} ) = -{\rm Im} ( V^*_{ub}V_{ud} V_{cb}V^*_{cd} )~,
\eeq
and in the U-spin limit we have
\beq
T_d = T_s ~~,~~~~ P_d = P_s ~.
\eeq
U-spin symmetry therefore leads to a relation between the $\btod$ and
$\btos$ decays:
\beq
| {\cal A}_d |^2 - | {\bar{\cal A}}_d |^2 = - \left[ | {\cal A}_s |^2 - | {\bar{\cal A}}_s |^2 \right] ~.
\label{Uspinreln}
\eeq

In general, there are four observables in the $\btod$ and $\btos$
processes: the branching ratios $B_d$ and $B_s$, and the direct CP
asymmetries $A^{CP}_d$ and $A^{CP}_s$. Eq.~(\ref{Uspinreln}) implies
that these are not independent, but obey \cite{Fleischer99,Uspinrel}
\beq
-\frac{A^{CP}_s}{A^{CP}_d} \, \frac{\tau(\bd) B_s}{\tau(\bs) B_d} = 1 ~.
\label{Uspinrelnobs}
\eeq
Thus, there are only three independent observables.

\subsection{\bf \boldmath $B_{d,s}^0 \to \ks h^+ h^-$ decays}
\label{UspinKhh}

We now turn to $B_{d,s}^0 \to \ks h^+ h^-$ decays. For definitiveness,
we focus on the pair ($\bs \to \ks \pi^+ \pi^-$ ($\btod$), $\bd \to
\ks K^+ K^-$ ($\btos$)), but the results can be equally applied to
($\bs \to \ks K^+ K^-$ ($\btod$), $\bd \to \ks \pi^+ \pi^-$
($\btos$)).

As discussed in Sec.~\ref{DP}, one must pay attention to the momenta
of the final-state particles. Let us define $f_d \equiv \ks(p_1)
\pi^+(p_2) \pi^-(p_3)$ and $\fbar_d \equiv \ks(p_1) \pi^+(p_3)
\pi^-(p_2)$, and similarly for $f_s$ and $\fbar_s$. Now consider
${\cal A}_d = {\cal A}(\bs \to f_d)$ and ${\cal A}_s = {\cal A}(\bd
\to f_s)$. The decay amplitudes ${\cal A}_d$ and ${\cal A}_s$ are
again given by Eqs.~(\ref{Addef}) and (\ref{Asdef}), respectively, and
are repeated below for convenience:
\beq
{\cal A}_d = V^*_{ub}V_{ud} T_d + V^*_{cb}V_{cd} P_d ~~,~~~~
{\cal A}_s = V^*_{ub}V_{us} T_s + V^*_{cb}V_{cs} P_s ~.
\label{3bodyproc}
\eeq
As these are three-body decays, $T_{d,s}$ and $P_{d,s}$ are all
momentum-dependent. This means that $T_d$ takes different values at
different points of the Dalitz plot, and similarly for $T_s$ and
$P_{d,s}$. For the CP-conjugate amplitudes, we have
\beq
{\bar{\cal A}}_d = V_{ub}V^*_{ud} {\bar T}_d + V_{cb}V^*_{cd} {\bar P}_d ~~,~~~~
{\bar{\cal A}}_s = V_{ub}V^*_{us} {\bar T}_s + V_{cb}V^*_{cs} {\bar P}_s ~.
\label{3bodyCPconj}
\eeq
Because the final states in the CP-conjugate decays are not the same
as in the decays ($p_2$ and $p_3$ are exchanged), $T_d \ne {\bar
  T}_d$, and similarly for $T_s$ and $P_{d,s}$.

We then have \cite{BGR}
\bea
| {\cal A}_d |^2 - | {\bar{\cal A}}_d |^2 & = &
              2 \, {\rm Im} ( V^*_{ub}V_{ud} V_{cb}V^*_{cd} ) \, {\rm Im} (T_d P_d^* + {\bar T}_d^* {\bar P}_d) ~, \nn\\
| {\cal A}_s |^2 - | {\bar{\cal A}}_s |^2 & = &
              2 \, {\rm Im} ( V^*_{ub}V_{us} V_{cb}V^*_{cs} ) \, {\rm Im} (T_s P_s^* + {\bar T}_s^* {\bar P}_s) ~.
\eea
In the U-spin limit we have $T_d = T_s$, $P_d = P_s$, ${\bar T}_d =
{\bar T}_s$, ${\bar P}_d = {\bar P}_s$, and the U-spin relation of
Eq.~(\ref{Uspinreln}) is reproduced.  However, since the amplitudes
themselves are now momentum dependent, this relation holds at each
point in the Dalitz plots.

As in the two-body case, the U-spin relation implies a relation among
the observables, similar to Eq.~(\ref{Uspinrelnobs}). This relation
involves $B^0 \to f$ and ${\bar B}^0 \to \fbar$ decays, and can be
written as
\beq
-\frac{a^{CP}_s}{a^{CP}_d} \, \frac{\tau(\bd) b_s}{\tau(\bs) b_d} = 1 ~.
\label{Uspinreln3body}
\eeq
Here, $a^{CP}_q$ and $b_q$ are, respectively, the direct CP asymmetry
and branching ratio defined locally, i.e., at a particular
Dalitz-plot point. They are both momentum-dependent quantities.

The analysis can be repeated for the case where ${\cal A}_d = {\cal
  A}(\bs \to \fbar_d)$ and ${\cal A}_s = {\cal A}(\bd \to \fbar_s)$.
Here we have
\beq
{\cal A}_d = V^*_{ub}V_{ud} {\bar T}_d + V^*_{cb}V_{cd} {\bar P}_d ~~,~~~~
{\cal A}_s = V^*_{ub}V_{us} {\bar T}_s + V^*_{cb}V_{cs} {\bar P}_s ~.
\label{3bodyproc2}
\eeq
and
\beq
{\bar{\cal A}}_d = V_{ub}V^*_{ud} T_d + V_{cb}V^*_{cd} P_d ~~,~~~~
{\bar{\cal A}}_s = V_{ub}V^*_{us} T_s + V_{cb}V^*_{cs} P_s ~.
\label{3bodyCPconj2}
\eeq
Once again, the U-spin relation of Eq.~(\ref{Uspinreln}) is
reproduced. And there is a relation like Eq.~(\ref{Uspinreln3body})
among the observables. This relation involves $B^0 \to \fbar$ and
${\bar B}^0 \to f$ decays.

The point here is that, for three-body decays, there are two U-spin
relations among the observables. These involve the same
momentum-dependent hadronic parameters.

\section{\boldmath Extraction of $\gamma$}
\label{extractgamma}

Here we present the details of how $\gamma$ can be extracted from a
U-spin analysis of $B_{d,s}^0 \to \ks h^+ h^-$ decays. We begin with a
review of the method for two-body decays.

\subsection{\bf Two-body decays}

The method proposed by Fleischer for extracting $\gamma$ from $\bskk$
and $\bdpipi$ \cite{Fleischer99} works as follows. The amplitude for
the $\btod$ decay ($\bdpipi$) is given in Eq.~(\ref{Addef}), which can
be written
\beq
{\cal A}_d = |V^*_{ub}V_{ud}| e^{i\gamma} T_d - |V^*_{cb}V_{cd}| P_d ~,
\eeq
where we have used $|V_{cd}| = - V_{cd}$. The amplitude for the
$\btos$ decay ($\bskk$) can be written similarly:
\beq
{\cal A}_s = |V^*_{ub}V_{us}| e^{i\gamma} T_s + |V^*_{cb}V_{cs}| P_s ~.
\eeq
In the U-spin limit, we have $T_d = T_s \equiv T$ and $P_d = P_s
\equiv P$.  Assuming that the magnitudes of the CKM matrix elements
are known, ${\cal A}_d$ and ${\cal A}_s$ each contain the same four
unknown parameters: $|T|$, $|P|$, their relative strong phase, and
$\gamma$.

Above [Eq.~(\ref{Uspinrelnobs})], it was noted that the branching
ratios and the direct CP asymmetries of these two decays are not
independent. Thus, $\gamma$ cannot be extracted from the measurements
of these observables alone, since there are more unknown theoretical
parameters (four) than observables (three). However, if the indirect
CP asymmetries in both $\bdpipi$ and $\bskk$ are also measured, and
values for the $\bd$-$\bdbar$ and $\bs$-$\bsbar$ mixing phases
($\beta$ and $\beta_s$, respectively) are taken from independent
measurements, there will be more observables (five) than unknowns,
which will allow $\gamma$ to be extracted.

\subsection{\bf \boldmath $B_{d,s}^0 \to \ks h^+ h^-$ decays}

A similar logic can be applied to three-body decays. However, care
must be taken in identifying the observables to be used, and in
establishing how these observables depend on the unknown theoretical
parameters.

The point is the following. If a final state $f$ is self-conjugate at
the quark level, both $B^0$ and ${\bar B}^0$ can decay to it. In the
case of two-body decays, the fact that $f$ is self-conjugate implies
that $\fbar = f$, so that the two decays $B^0,{\bar B}^0 \to f$ must
be considered.  However, as noted in Sec.~\ref{DP}, for three-body
decays, a self-conjugate $f$ still has $\fbar \ne f$, since $f$ and
$\fbar$ correspond to different points of the Dalitz plot. In this
case, the analysis must consider the four decays $B^0,{\bar B}^0 \to
f, \fbar$. The time dependence of two-body decays has been analyzed in
Refs.~\cite{DR,Dunietzwidth}. Below we adapt this analysis to
three-body decays.

In the presence of $B^0$-${\bar B}^0$ mixing, the $B_L$ and $B_H$
states ($L$ is light, $H$ is heavy) are mixtures of $B^0$ and $\ob$.
The physical time-dependent neutral $B$-meson states can then be
expressed as
\bea
\ket{B^0_\phys(t)} &=& f_+(t)\ket{B^0} + \frac{q}{p}f_-(t)\ket{\ob}
~,~~ \nn\\
\ket{\ob_\phys(t)} &=& \frac{p}{q}f_-(t)\ket{B^0} + f_+(t)\ket{\ob}
~.
\eea
Here $B^0_\phys(t)$ ($\ob_\phys(t)$) is the state that is a $B^0$
($\ob$) at $t=0$. In the above, $q/p = e^{-2i\phi_M}$, where $\phi_M$
is the weak phase of the mixing (the $\bd$-$\bdbar$ and $\bs$-$\bsbar$
mixing phases are $\beta$ and $\beta_s$, respectively), and
\beq
f_+(t) = e^{-i(m - i\Ga/2)t}\cos(\De\mu t/2) ~~,~~~~
f_-(t) = e^{-i(m - i\Ga/2)t}i\sin(\De\mu t/2)~,
\eeq
with
\bea
& m = (m_H + m_L)/2 ~~,~~~~ \De m = m_H - m_L ~, & \nn\\
& \Ga = (\Ga_H + \Ga_L)/2 ~~,~~~~ \De\Ga = \Ga_H - \Ga_L ~, & \nn\\
& \De\mu = \De m - i\De\Ga/2 ~. &
\eea
The decay amplitudes are then given by
\bea
\brkt{f}{B^0_\phys(t)} &=& \brkt{f}{B^0}\(f_+(t) + \la f_-(t)\)
~,~~ \nl
\brkt{\of}{B^0_\phys(t)} &=& \frac{q}{p}\brkt{\of}{\ob}\(f_+(t)
{\bar\la} + f_-(t)\) ~,~~ \nl
\brkt{f}{\ob_\phys(t)} &=& \frac{p}{q}\brkt{f}{B^0}\(f_-(t)
+ \la f_+(t)\) ~,~~ \nl
\brkt{\of}{\ob_\phys(t)} &=& \brkt{\of}{\ob}\(f_-(t)\ola
+ f_+(t)\) ~,
\eea
where
\bea
x \equiv \frac{\brkt{f}{\ob}}{\brkt{f}{B^0}} ~,~~
\ox \equiv \frac{\brkt{\of}{B^0}}{\brkt{\of}{\ob}}~,~~
\la \equiv \frac{q}{p} x ~,~~
\ola \equiv \frac{p}{q}{\bar x} ~.
\eea

In Ref.~\cite{DR} the assumption is made that $\De\Ga = 0$. In
Ref.~\cite{Dunietzwidth} it is noted that $\De\Ga$ is nonzero in $\bs$
decays. Our expressions below therefore allow for a nonzero $\De\Ga$.

The decay rates are proportional to the squares of the amplitudes,
which take the form
\bea
|\cM|^2(B^0_\phys(t)\to f) &=& \frac12 |A|^2 e^{-\Ga t} \Bigl[
(1 - |x|^2)\cos(\De m t) + (1 + |x|^2)\cosh(\De\Ga t/2) \nl
&&\hskip1.5truecm  - 2 \, \Im(\la)\sin(\De m t) + 2 \, \Re(\la)\sinh(\De\Ga t/2) \Bigr] ~, \nl
|\cM|^2(\ob_\phys(t)\to f) &=& \frac12 |A|^2 e^{-\Ga t} \Bigl[
-(1 - |x|^2)\cos(\De m t) + (1 + |x|^2)\cosh(\De\Ga t/2) \nl
&&\hskip1.5truecm  + 2 \, \Im(\la)\sin(\De m t) + 2 \, \Re(\la)\sinh(\De\Ga t/2) \Bigr] ~, \nl
|\cM|^2(B^0_\phys(t)\to\of) &=& \frac12 |\oA|^2 e^{-\Ga t} \Bigl[
-(1 - |\ox|^2)\cos(\De m t) + (1 + |\ox|^2)\cosh(\De\Ga t/2) \nl
&&\hskip1.5truecm  + 2 \, \Im(\ola)\sin(\De m t) + 2 \, \Re(\ola)\sinh(\De\Ga t/2) \Bigr] ~, \nl
|\cM|^2(\ob_\phys(t)\to\of) &=& \frac12 |\oA|^2 e^{-\Ga t} \Bigl[
(1 - |\ox|^2)\cos(\De m t) + (1 + |\ox|^2)\cosh(\De\Ga t/2) \nl
&&\hskip1.5truecm  - 2 \, \Im(\ola)\sin(\De m t) + 2 \, \Re(\ola)\sinh(\De\Ga t/2) \Bigr] ~,
\label{ampssquared}
\eea
where $A \equiv \brkt{f}{B^0}$, $\oA \equiv \brkt{\of} {\ob}$, and we
have used $|q/p| = 1$.

With the squares of the amplitudes in hand, we can now obtain
expressions for the observables. Before doing so, there is one point
that must be mentioned. Although we have referred to measurements at
different points of the Dalitz plot, in practice it is only possible
to make measurements in bins, i.e., over areas of the Dalitz plot
centred at different points. The observables will then involve
integrals over the Mandelstam variables representing these bins.

Using the first two equations of Eq.~(\ref{ampssquared}), we can now
construct the time-dependent CP-averaged rate and the CP asymmetry for
the final state $f$:
\bea
\label{obsBR}
\Ga(t) &=& \hf(\Ga(B^0_\phys(t)\to f) + \Ga(\ob_\phys(t)\to f))~,~~ \nl
&=& \hf\iint\limits_{\bin} ds_{12} ds_{23} \, |A|^2e^{-\Ga t}\[\(1 + |x|^2\)\cosh(\De
\Ga t/2)\r.~~~\nl
&&\hskip5truecm \l.+~ 2 \Re(\la)\sinh(\De\Ga t/2))\] ~,~~ \\
\label{obsACP}
A_{CP}(t) &=& \frac{\Ga(B^0_\phys(t)\to f) - \Ga(\ob_\phys(t)\to f)}
{\Ga(B^0_\phys(t)\to f) + \Ga(\ob_\phys(t)\to f)} ~,~~ \nl
&=& \frac{\iint\limits_{\bin} ds_{12} ds_{23} \,  |A|^2\[(1 - |x|^2)\cos(\De m t)
- 2\Im(\la)\sin(\De m t)\]}{\iint\limits_{\bin} ds_{12} ds_{23} \, |A|^2
\[(1 + |x|^2)\cosh(\De\Ga t/2) + 2 \Re(\la)\sinh(\De\Ga t/2)\]} ~.~~
\eea
In $\Ga(t)$, one does not distinguish $B^0_\phys(t)$ and
$\ob_\phys(t)$ decays, whereas one does in $A_{CP}(t)$. Thus, as
usual, the measurement of the CP asymmetry requires tagging.

A comment should be made about Eq.~(\ref{obsACP}). The direct CP
asymmetry compares $B^0 \to f$ and ${\bar B}^0 \to \fbar$ decays.
Because $\fbar = f$ in two-body decays, there one refers to the
coefficient of $\cos(\De m t)$ as the direct CP asymmetry.  However,
in three-body decays, because $\fbar \ne f$, the situation is
different. Here the coefficient of $\cos(\De m t)$ compares $B^0 \to
f$ and ${\bar B}^0 \to f$ decays, and so it is not actually a CP
asymmetry.

In the above definitions there appear to be four observables, namely
the coefficients of $\cos(\De m t), \cosh(\De\Ga t/2), \sin(\De m t)$,
and $\sinh(\De\Ga t/2)$, as can be determined from $\Ga(t)$ and the
numerator of $A_{CP}(t)$. However, these coefficients are not all
independent, as can be seen in the following identity:
\bea
|A|^2(1 + |x|^2) - |A|^2(1 - |x|^2) &=& 2|A|^2|x|^2 ~=~ 2|A|^2|\la|^2 ~,~~\nl
&=& 2|A|^2\(\Re(\la)^2 + \Im(\la)^2\) ~.~~
\eea
There are therefore only three independent observables.

One can perform a similar analysis using the last two equations of
Eq.~(\ref{ampssquared}). In this way one constructs the time-dependent
CP-averaged rate and the CP asymmetry for the final state $\fbar$.
There are again three independent observables. Thus, for a given
$B_{d,s}^0 \to \ks h^+ h^-$ decay, there are a total of six
observables: three each for the final states $f$ and $\fbar$.

We now turn to the question of the number of unknown theoretical
parameters, focusing on the decay pair $\bs \to \ks \pi^+ \pi^-$
($\btod$) and $\bd \to \ks K^+ K^-$ ($\btos$). Consider first the
$\btod$ decay.  The amplitudes for the various $B^0,{\bar B}^0 \to f,
\fbar$ decays are given in Eqs.~(\ref{3bodyproc}),
(\ref{3bodyCPconj}), (\ref{3bodyproc2}) and (\ref{3bodyCPconj2}).
There are eight unknown parameters: $|T_d|$, $|P_d|$, $|{\bar T}_d|$,
$|{\bar P}_d|$, their three relative strong phases, and $\gamma$. With
only six observables, $\gamma$ cannot be extracted.

This can be remedied by also considering the U-spin conjugate $\btos$
decay $\bd \to \ks K^+ K^-$. Its $B^0,{\bar B}^0 \to f, \fbar$
amplitudes are also given in Eqs.~(\ref{3bodyproc}),
(\ref{3bodyCPconj}), (\ref{3bodyproc2}) and (\ref{3bodyCPconj2}).
Here too there are eight unknown parameters: $|T_s|$, $|P_s|$, $|{\bar
  T}_s|$, $|{\bar P}_s|$, their three relative strong phases, and
$\gamma$. However, in the U-spin limit, we have $T^{}_d = T^{}_s
\equiv T^{}$ and $P^{}_d = P^{}_s \equiv P^{}$. Thus, the two decays
are described by the same eight unknown parameters.  (As before, it is
assumed that the $\bd$-$\bdbar$ and $\bs$-$\bsbar$ mixing phases are
taken from independent measurements.) But there are now twelve
observables, six for each of $\bs \to \ks \pi^+ \pi^-$ and $\bd \to
\ks K^+ K^-$. On the other hand, it was noted in Sec.~\ref{UspinKhh}
that there are two U-spin relation among the branching ratios and
direct CP asymmetries of the $\btod$ and $\btos$ decays. Still, this
leaves ten independent observables, which is more than the number of
unknown parameters. Thus, assuming again that the magnitudes of the
CKM matrix elements are known, $\gamma$ can be extracted.

It must be mentioned that this method introduces a new systematic
error. We have argued above that since the number of observables is
greater than the number of unknowns, $\gamma$ can be extracted. But
this only works if all the observables are functions of the same
unknowns. And because the measurements must be made using bins of the
Dalitz plot, this does not hold exactly. Writing $A~x = \brkt{f}{\ob}
= \tA$, from Eqs.\ (\ref{obsBR}) and (\ref{obsACP}) we have
\bea
BR &\propto& \iint\limits_{\bin} ds_{12} ds_{23} \,  (|A|^2 + |\tA|^2) ~, \nl
A^{CP}_{dir} &\propto& \iint\limits_{\bin} ds_{12} ds_{23} \,  (|A|^2 - |\tA|^2) ~, \nl
A^{CP}_{indir} &\propto& \iint\limits_{\bin} ds_{12} ds_{23} \,  \Im[ (q/p) A^* \tA ] ~.
\eea
If we define
\beq
\iint\limits_{\bin} ds_{12} ds_{23} \,  |A|^2 \equiv |A'|^2 ~~,~~~~
\iint\limits_{\bin} ds_{12} ds_{23} \,  |\tA|^2 \equiv |\tA'|^2 ~,
\eeq
we see that both $BR$ and $A^{CP}_{dir}$ are functions of $A'$ and
$\tA'$. However, $A^{CP}_{indir}$ is not. We must make the
approximation that
\beq
\iint\limits_{\bin} ds_{12} ds_{23} \,  \Im[ (q/p) A^* \tA ] \simeq \Im[ (q/p) {A'}^*
\tA' ] ~,
\eeq
and this introduces a systematic error. The above holds exactly for a
single point of the Dalitz plot. Thus, the smaller the bins are, the
better is the approximation, leading to a smaller systematic error. On
the other hand, smaller bins lead to larger statistical errors. The
bin size must therefore be chosen to minimize the total error.

\section{U-Spin Breaking}

As noted earlier, the method of combining measurements of decays
related by U spin to extract $\gamma$ was originally proposed in the
context of two-body decays \cite{Fleischer99}. Here, there is a
theoretical error due to unknown U-spin-breaking effects. This same
difficulty arises when applying the method to three-body decays.  In
this section we examine the question of U-spin breaking as pertains to
three-body decays.

The method described in the previous section for extracting $\gamma$
involves combining measurements of pairs of three-body decays related
by U-spin, such as $\bs \to \ks \pi^+ \pi^-$ ($\btod$) and $\bd \to
\ks K^+ K^-$ ($\btos$). This method applies at a particular pair of
Dalitz-plot points (bins). By repeating this analysis for all points,
this provides multiple measurements of $\gamma$. These can then be
averaged over the entire Dalitz plot, reducing the statistical error.

In the presence of U-spin breaking, the extracted value of $\gamma$,
$\gamma_{ext}$, will differ from its true value, $\gamma_{true}$. Now,
there are several different U-spin-breaking parameters. However, these
parameters are all momentum dependent. Thus, their effect on the
extracted value of $\gamma$ will vary from point to point on the
Dalitz plot. That is, if
\beq
\gamma_{ext} - \gamma_{true} = N ~,
\label{gammaN}
\eeq
it is likely that $N > 0$ at some points, and $N < 0$ at others. In
this case, averaging over all Dalitz-plot points will also reduce the
effect of U-spin breaking, so that $(\gamma_{ext})_{avg}$ will
approach $\gamma_{true}$. If this occurs, the main theoretical error
of the method will be significantly reduced.

Still, while this is a nice idea, how can we be certain that it is
happening? Fortunately, there is a way of experimentally testing
whether or not this behaviour is present in three-body decays. In
Eq.~(\ref{Uspinreln3body}) it was shown that there is a relation among
the observables of two decays related by U-spin reflection.  Writing
\beq
-\frac{a^{CP}_s}{a^{CP}_d} \, \frac{\tau(\bd) b_s}{\tau(\bs) b_d} -
1 = n' ~,
\label{Uspinreln3bn'}
\eeq
we have $n'=0$ in the U-spin limit. By measuring $b_{d,s}$ and
$a^{CP}_{d,s}$, and constructing the above ratio at each Dalitz-plot
point, it is possible to experimentally determine if an average over
all points leads to $n' \to 0$.

The above test requires a Dalitz analysis. A simpler test of U-spin
breaking can be obtained by separately integrating the numerator and
denominator of Eq.~(\ref{Uspinreln3bn'}) over the
kinematically-allowed regions of the Dalitz plots (denoted by $\DP$):
\beq
-\frac{\tau(\bd)}{\tau(\bs)} \, \frac{\iint\limits_{\DP} ds_{12}ds_{23}a^{CP}_s b_s}
{\iint\limits_{\DP} ds_{12}ds_{23}a^{CP}_d b_d} - 1 = -\frac{A^{CP}_s}{A^{CP}_d} \,
\frac{\tau(\bd) B_s}{\tau(\bs) B_d} - 1 = N' ~.
\label{Uspinreln3bintN'}
\eeq
Unlike $n'$, which is defined using momentum-dependent quantities,
$N'$ depends only on integrated quantities, and hence does not depend
on final-state momenta. Once again, we have $N' = 0$ in the U-spin
limit.

The above tests can be carried out using the measurements of
$B_{d,s}^0 \to \ks h^+ h^-$ decays. However, it is not necessary to
wait until these are made. Other pairs of three-body decays related by
U spin are (i) $B^+ \to \pi^+ K^+ K^-$ ($\btod$) and $B^+ \to K^+
\pi^+ \pi^-$ ($\btos$), and (ii) $B^+ \to \pi^+ \pi^+ \pi^-$ ($\btod$)
and $B^+ \to K^+ K^+ K^-$ ($\btos$). In Ref.~\cite{Uspinbreaking},
group theory is used to write the factor $n'$ of
Eq. (\ref{Uspinreln3bn'}) for these decay pairs in terms of
U-spin-breaking parameters. These parameters take into account all
U-spin-breaking effects, such as differences in the masses of the
$\pi$ and $K$ mesons, differences in the properties of the resonances
contributing to the decays (e.g., $\rho$, $\phi$), etc. It is found
that, to first order, $n'$ is proportional to a linear combination of
such parameters. That is, in the presence of U-spin breaking, $n'\ne
0$. However, the U-spin-breaking parameters are
momentum-dependent. Using the same logic as before, it would not be
surprising to find $n' > 0$ at some points and $n' < 0$ at others.  If
so, the average over all Dalitz-plot points will reduce the effect of
U-spin breaking in the above relation.

These $B^+$ decays have recently been measured by LHCb
\cite{LHCbDeltaS=1, LHCbDeltaS=0}. In Ref.~\cite{BGR}, the U-spin
relation of Eq.~(\ref{Uspinreln3bintN'}) is tested using data
integrated over the Dalitz plot. We have updated these results with
more recent data from Refs.~\cite{LHCUpdate}. The updated
results are shown in Table \ref{tab:1}. Unfortunately, at present the
results are simply not precise enough to draw any conclusions. When
the data improve, we will have a better idea of whether averaging (or
integrating) over the Dalitz plot reduces the effect of U-spin
breaking.

\begin{table}[htb]
\center
\begin{tabular}{ccc}
\hline
\hline
Asymmetry ratio & U-spin & LHCb \\
& prediction & result \\
\hline
$A^{CP}(B^+ \to \pi^+ K^+ K^-)/A^{CP}(B^+ \to K^+ \pi^+ \pi^-)$ & $-10.2 \pm 1.5$ & $-4.9 \pm 2.0$ \\
$A^{CP}(B^+ \to \pi^+ \pi^+ \pi^-)/A^{CP}(B^+ \to K^+ K^+ K^-)$ & $-2.2 \pm 0.2$ & $-1.6 \pm 0.5$ \\
\hline
\hline
\end{tabular}
\caption{U-spin predictions for asymmetry ratios compared with LHCb
  measurements.}
\label{tab:1}
\end{table}

Finally, another source of U-spin breaking arises from the fact that
$\pi^\pm$ and $K^\pm$ do not have the same mass, and similarly for
$\bd$ and $\bs$. This results in a difference between the
kinematically-allowed phase space for a decay and that for its U-spin
partner. Due to this difference, there will be regions of the Dalitz
plots where the observables defined in Sec.~\ref{extractgamma} can be
obtained only for one of the two decays being compared. These regions
must be excluded from the analysis, since our method for extracting
$\gamma$ works only for those regions of the Dalitz plots where the
two decays have overlapping kinematically-allowed regions.

\section{Conclusions}

In 1999, R. Fleischer proposed a method for extracting $\gamma$ using
a pair of two-body decays whose amplitudes are related by U-spin
symmetry ($d\leftrightarrow s$) \cite{Fleischer99}. It involves
combining the measurements of the branching ratios and CP asymmetries,
both direct and indirect, of the two decays. These decay amplitudes
include penguin diagrams, which may receive important (loop-level)
contributions from new physics. If so, the value of $\gamma$ extracted
using this method will disagree with its current value, which is
obtained using tree-level decays.

In the present paper we adapt this method to charmless $B \to PPP$
decays ($P$ is a pseudoscalar meson), specifically $B^0_{d,s}\to K_S
h^+ h^-$ ($h = K,\pi$). Time-dependent Dalitz analyses of the
three-body decays can be used to measure the branching fractions and
CP asymmetries. Note that it is not necessary to perform an isobar
amplitude analysis of the Dalitz plot. We show that there are more
observables than unknown theoretical parameters, so that $\gamma$ can
be extracted by fitting to the observables. The decay amplitudes for
three-body decays depend on the momenta of the final-state
particles. The method applies to each point of the Dalitz plot, and
thus constitutes many independent measurements of $\gamma$.

The main source of theoretical error in the extraction of $\gamma$,
which also applies to the method with two-body decays, is U-spin
breaking. However, three-body decays offer the potential to reduce
this error. The U-spin-breaking effects are also momentum-dependent.
As such, the difference between the extracted value of $\gamma$ and
its true value may well vary, in both magnitude and sign, from point
to point in the Dalitz plot. If this is the case, then averaging over
the Dalitz plot will reduce the error due to U-spin breaking.

It is possible to test experimentally whether or not this behaviour is
present in three-body decays. In the U-spin limit, there is a relation
among the branching ratios and direct CP asymmetries of the two decays
that are related by U spin. This applies to the decays $B^+ \to \pi^+
K^+ K^-$ ($\btod$) and $B^+ \to K^+ \pi^+ \pi^-$ ($\btos$), and $B^+
\to \pi^+ \pi^+ \pi^-$ ($\btod$) and $B^+ \to K^+ K^+ K^-$ ($\btos$),
all of which have been measured. Unfortunately, the current data on
these decays still has large errors, so that it is unclear whether
U-spin breaking is small when averaged over the Dalitz plot. Future
precision data in these channels will be able to clearly show the
size of this U-spin breaking.

\bigskip
\noindent
{\bf Acknowledgments}: We thank M. Imbeault, T. Gershon, M. Gronau and
J. Rosner for helpful conversations and communications.  This work was
financially supported by the IPP (BB) and by NSERC of Canada (BB,DL).


\end{document}